\documentclass[twocolumn]{revtex4}

\usepackage{float}
\usepackage{subfigure}
\usepackage{amsmath,amssymb}
\usepackage{graphicx}
\usepackage{dcolumn}
\usepackage{bm}
\usepackage{lineno}

\newcommand{\cool}{(5S_{1/2} \ F = 2)\rightarrow (5P_{3/2}\  F^\prime = 3)}

\newcommand{\figref}[1]{Fig.~\ref{#1}}
\begin{document}

\title{Temperature and phase-space density of cold atom cloud in a quadrupole magnetic trap}
\author{S. P. Ram}
\email{spram@rrcat.gov.in}
\author{S. R. Mishra}
\author{S. K. Tiwari}
\author{H. S. Rawat}

\affiliation{
Laser Physics Applications Section,\\
Raja Ramanna Centre for Advanced Technology,\\
Indore 452013 India. \\
}%


\begin{abstract}
We present studies on the modifications in temperature, number density and phase-space density when a laser cooled atom cloud from the optical molasses is trapped in a quadrupole magnetic trap. Theoretically it is shown that for a given temperature and size of the cloud from the molasses, the phase-space density in the magnetic trap first increases with magnetic field gradient and then decreases with it, after attaining a maximum value at an optimum value of magnetic field gradient. The experimentally measured variation in phase-space density in the magnetic trap with the magnetic field gradient has shown the similar trend. However, the experimentally measured values of number density and phase-space density are much lower than their theoretically predicted values. This is attributed to the higher experimentally observed temperature in the magnetic trap than the theoretically predicted temperature. Nevertheless, these studies can be useful to set a higher phase-space density in the trap by setting the optimum value of field gradient of the quadrupole magnetic trap.
\end{abstract}


\maketitle
\section{Introduction}
\label{sec:introduction}
The first demonstration of magnetic trapping of neutral atoms \cite{Migdall1985a} proved to be an instrumental idea in achieving the Bose-Einstein condensation (BEC) of Rubidium atoms in a magnetic trap after evaporative cooling \cite{Anderson1995a,Bradley1995a,Davis1995a,Inguscio1999a}. The trapping and cooling of atoms to very low temperature has various applications in atom interferometry based precision measurements (of time, acceleration, rotation and force), atom lithography, quantum information etc. {\cite{Borde1991a,Borde1997a,Peters2001a,Arimondo2005a,Garcia-Ripoll2005a,Perrin2011a}. For the realization of BEC in neutral atoms, the laser cooling of atoms is the first stage of cooling which is followed by the second stage cooling, i.e. evaporative cooling. For evaporative cooling, various traps such as magnetic trap \cite{Petrich1994a, Petrich1995a}, optical dipole trap \cite{Barrett2001a,Arnold2011a}, or a hybrid of magnetic and optical trap \cite{Lin2009a} are commonly used. For trapping of cold neutral atoms several designs of magnetic traps have been proposed and demonstrated \cite{Bergeman1987a,Inguscio1999a,Xu2001b,Yin2002a}. In the magnetic trap approach, the higher energy atoms from the trap are evaporated by the application of radio frequency radiation, leading to decrease in the temperature and increase in the in phase-space density. The phase-space density ($\rho$) is given as $\rho~=~n\lambda_{dB}^3$, where $n$ is the number density and $\lambda_{dB}$ is the thermal de-Broglie wavelength for the atom. After the evaporative cooling, the final phase-space density required to achieve BEC must satisfy the condition $\rho>1.202$ (in a harmonic trap) \cite{Pitaevskii2003a}. Thus the initial phase-space density of atoms in the magnetic trap is of significant importance which depends upon various parameters during the transfer of laser cooled atom cloud from the molasses to quadrupole magnetic trap. To obtain the maximum value of $\rho$ in the magnetic trap after the atom cloud is transferred from molasses to the magnetic trap, the value of parameters such as temperature and size of the atom cloud in the molasses govern the optimum magnetic field gradient of the trap.

In this work, we have studied the changes in the phase-space density of laser cooled atom cloud after its transfer from molasses to a quadrupole magnetic trap. Theoretically, it is shown that for a given temperature and size of the laser cooled atom cloud, there is an optimum value of magnetic field gradient of the quadrupole trap to obtain the maximum phase-space density of atom cloud in the magnetic trap after the capture of laser cooled atom cloud in the magnetic trap. The experimentally measured number density and phase-space density of atoms in the magnetic trap have shown the similar trend in variation with the magnetic field gradient of the trap. However, the experimentally measured phase-space density values are much lower than the theoretically estimated values. This is attributed to the higher values of measured temperature in the magnetic trap than the theoretically predicted values of temperature. This study is useful to set the higher initial phase-space density of the cloud in the quadrupole magnetic trap before the evaporative cooling. Since the adiabatic compression of the quadrupole trap conserves the phase-space density ($\rho$), the optimized value of $\rho$ obtained during the quadrupole trap loading (by switching-on the optimum field gradient) can be preserved during the adiabatic compression \cite{Pinkse1997a}. The higher initial phase-space density in the quadrupole trap can be useful when this quadrupole trap is transformed into a hybrid trap \cite{Lin2009a} or quadrupole Ioffe configuration (QUIC) trap \cite{Esslinger1998a} to perform evaporative cooling experiments.

\section{Theory}
\label{sec:theory}

\begin{figure}[!h]%
\centering
\includegraphics[width=0.9\columnwidth]{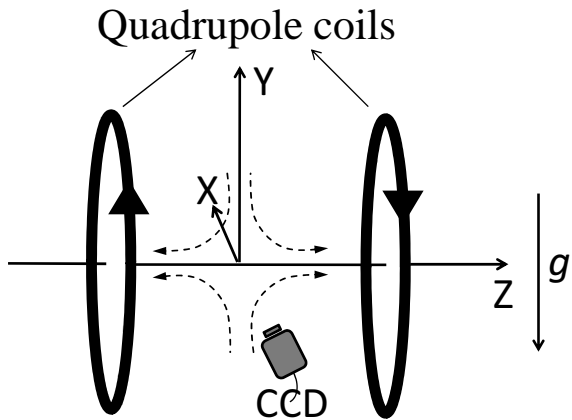}%
\caption{Schematic of the arrangement of the coils for quadrupole magnetic trap and axes of the coordinate system. The CCD is placed at the x-axis facing the trapped atom cloud as shown in this figure.}%
\label{fig:quad_trap_schematic}%
\end{figure}

We first theoretically calculate the phase-space density of the cold atom cloud ($^{87}$Rb atoms) after it is transferred from the molasses to a quadrupole magnetic trap. The atom cloud in the molasses is assumed to have $N$ number of atoms at a temperature $T_i$ and size $\sigma_i$ (i.e. root mean square (r.m.s.) radius of the number density profile of the cloud). The phase-space density of this initial cloud after molasses is given as \cite{Inguscio1999a}, 
\begin{equation}  
\rho_{i}=n_{i}\left[\frac{h^2}{2\pi mk_B T_i}\right]^{3/2}
\label{eq:initial_PSD}
\end{equation}
where $n_i$ is the number density in the molasses, $m$ is the mass of the atom, $h$ is the Planck's constant and $k_B$ is the Boltzmann's constant. The number density of atoms in the atom cloud in the molasses is assumed to be as
\begin{equation}
n_{i}(x,y,z)=n_i(0)\exp\left(-\frac{(x^2+y^2+z^2)}{2\sigma_i^2}\right),
\label{eq:initial_ND}
\end{equation}
which results in relation between the total number of atoms $N$ and and peak number density $n_i(0)$ in the molasses as
\begin{equation}
N=\int^{+\infty}_{-\infty}\int^{+\infty}_{-\infty}\int^{+\infty}_{-\infty}{n_i (x,y,z).dxdydz} =n_i (0)(2\pi\sigma_i^2 )^{3/2}.
\label{eq:total_number}
\end{equation}

Experimentally, it is standard practice that after cooling of atoms in the molasses, the cooling laser beams are switched-off and atoms are optically pumped to a trappable state before the capture of atoms in the magnetic trap. It is assumed that during the optical pumping, the rise in the temperature of the atom cloud is negligible. But, when the atom cloud is captured in the quadrupole magnetic trap, both, temperature and number density in the trapped cloud get modified due to interaction of atoms with the field of the trap. This finally leads to the modified temperature, number density and phase-space density ($\rho$) of the cloud in the magnetic trap as compared to those in the initial cloud obtained from the molasses. These modified values of the atom cloud parameters in the magnetic trap can be calculated as follows.

We consider a quadrupole magnetic trap whose configuration is as shown in \figref{fig:quad_trap_schematic}, where gravitational field direction is along the opposite direction of y-axis. The magnetic field near the center of this magnetic trap can be approximately written as $B=b[-x/2, -y/2, z]$ (as per current direction in \figref{fig:quad_trap_schematic}), where $b$ is the field gradient in axial direction ($z$- direction). When this magnetic trap is switched-on instantaneously, the atom cloud from molasses (which is assumed to be in the trappable state ($\left|F, ~m_F\right\rangle$)) gains a potential energy from the magnetic trap which is given as \cite{Mewes1997b,Landini2012a}, 
\begin{widetext}
\begin{equation}
PE=\left\langle -\vec{\mu}.\vec{B}\right\rangle=g_Fm_F\mu_Bb\int^{+\infty}_{-\infty}\int^{+\infty}_{-\infty}\int^{+\infty}_{-\infty}\left(\sqrt{\frac{x^2+y^2}{4}+z^2}\right)n_{i}(x,y,z)dxdydz,
\label{eq:PE1}
\end{equation}
\end{widetext}
where $m_F$ is magnetic hyperfine angular momentum quantum number, $g_F$ is Land\'e $g$-factor and $\mu_B$ is the Bohr magneton. On using $n_i$ from equation \eqref{eq:initial_ND} in equation \eqref{eq:PE1} and integrating over variables $(x, y, z)$, we can find the potential energy as
\begin{equation}
PE=17.34\frac{Ng_F m_F \mu_B b\sigma_i}{(2\pi)^{3/2}}.
\label{eq:PE}
\end{equation}

After the molasses and optical pumping, once the cold atom cloud is trapped in the magnetic trap and has reached the equilibrium with it, the total energy of the atoms in the magnetic trap is the sum of the initial kinetic energy and the potential energy gained from the magnetic trap (expressed in equations \eqref{eq:PE1} and \eqref{eq:PE}). In such an equilibrium state formed in the trap, the kinetic energy of the atoms is one third of the total energy of the atom cloud in quadrupole trap, as predicted by virial theorem \cite{Stuhler2001a}. This gives the temperature ($T_f$) of the atom cloud in the trap as, 
\begin{equation}
\frac{3}{2}Nk_BT_f=\frac{1}{3}\left[\frac{3}{2}Nk_BT_i+PE\right],
\label{eq:virial_theorem}
\end{equation}
which gives
\begin{equation}
T_f=\frac{T_i}{3}+\kappa \frac{\mu_B b\sigma_i}{k_B},
\label{eq:Tf}
\end{equation}
where $\kappa= 0.24$. The equation (7) is obtained after using $g_Fm_F =1$ for state $\left|F=2, m_F =2\right\rangle$ of $^{87}$Rb atoms. Assuming that all the atoms in the molasses are captured in the magnetic trap, the final peak number density $n_f (0)$ in the magnetic trap can be related to the number of atoms in the magnetic trap as, 
\begin{widetext}
\begin{equation}
N=n_f (0)\int^{+\infty}_{-\infty}\int^{+\infty}_{-\infty}\int^{+\infty}_{-\infty}\exp\left(-{\frac{g_Fm_F\mu_Bb\left(\sqrt{\frac{x^2+y^2}{4}+z^2}\right)+mgy}{k_BT_f}}\right)dxdydz.
\label{eq:N_mag_trap}
\end{equation}
\end{widetext}

For $(g_fm_f\mu_B b/2)>mg$, the above equation is analytically integrable to give rise to the expression of the final peak number density for state $\left|F=2, m_F =2\right\rangle$ of $^{87}$Rb atoms as
\begin{equation}
n_f (0)=\frac{N}{32\pi}\left(\frac{\mu_Bb}{\frac{k_BT_i}{3}+\kappa\mu_Bb\sigma_i}\right)^3\left[\left(\frac{2mg}{\mu_Bb}\right)^2-1\right]^2.
\label{eq:final_ND1}
\end{equation} 
\begin{figure}[!h]%
\centering
\includegraphics[width=0.9\columnwidth]{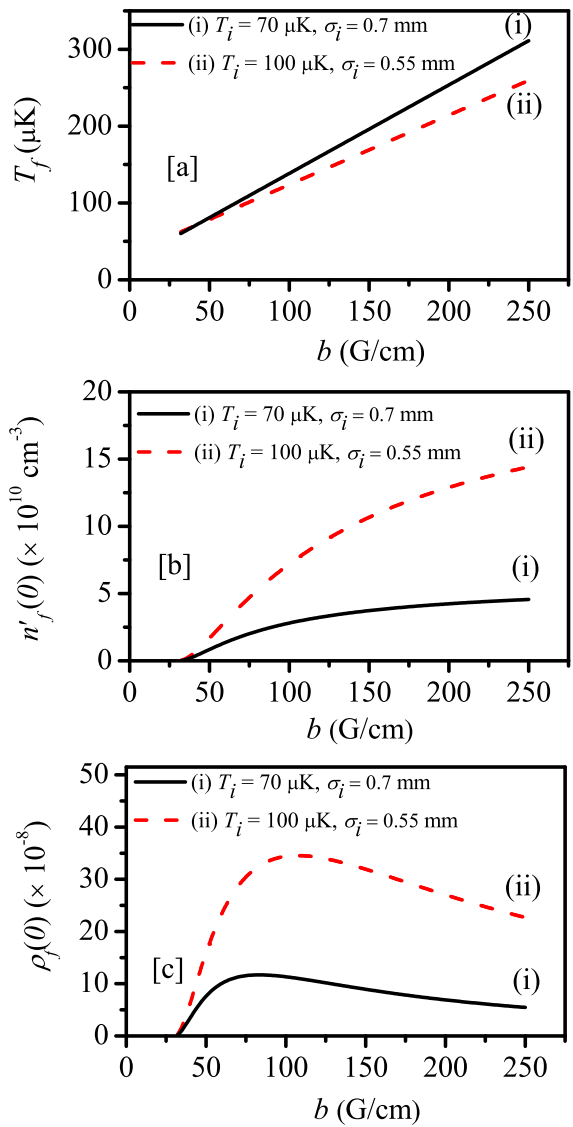}
\caption{The calculated variation in final temperature $T_f$ (figure [a]), final peak density $n^\prime_f(0)$ (figure [b]) and final phase-space density $\rho_f(0)$ (figure [c]) in the quadrupole magnetic trap with field gradient $b$ for the different values of temperature and size of the cloud after molasses. For graphs (i) and (ii) (in figures [b] and [c]), we used $N^\prime=3.0\times10^7$ and $5.5\times10^7$, which are close to the experimentally observed values of number of atoms.
}%
\label{fig:theory}%
\end{figure}

During the actual experiments, due to finite switching-on time of current in the magnetic trap coils, a fraction of the number of atoms in the cloud from the molasses may be lost due to fall under gravity as well as due to expansion of the cloud. The inefficient optical pumping also contributes to the reduction in number of atoms in the magnetic trap. Thus the number of atoms actually trapped in the magnetic trap ($N^\prime$) is always less than the number of atoms in the molasses ($N$). Considering these different loss processes, and defining `$\varepsilon$' as overall efficiency for the transfer of atoms from molasses to the magnetic trap, one can relate $N^\prime$ and $N$ as $N ^\prime = \varepsilon N$. The peak number density in the magnetic trap can, then, be re-written as
\begin{equation}
n_f^\prime (0)=\frac{N^\prime}{32\pi} \left(\frac{\mu_Bb}{\frac{k_BT_i}{3}+\kappa\mu_Bb\sigma_i}\right)^3\left[\left(\frac{2mg}{\mu_Bb}\right)^2-1\right]^2.                                    
\label{eq:final_ND}
\end{equation}

It is important to note here that the number density becomes independent of gravity when $b>>(2 m g/\mu_B) (i.e. \sim 31 ~G/cm ~ \mbox{for} ~^{87}\mbox{Rb})$ . Using the final peak density given by equation \eqref{eq:final_ND} and the final temperature $T_f$ in the magnetic trap given by equation \eqref{eq:Tf}, we can find the final peak phase-space density of the atom cloud in the quadrupole magnetic trap as,
\begin{widetext}
\begin{equation}
\rho_f(0)=n_f^\prime(0)\left[\frac{h^2}{2\pi mk_BT_f}\right]^{3/2}=\frac{N^\prime(\mu_Bbh)^3}{64\sqrt{2}\pi^{5/2}m^{3/2}}\frac{\left[\left(\frac{2mg}{\mu_Bb}\right)^2-1\right]^2}{\left(\frac{k_BT_i}{3}+\kappa\mu_Bb\sigma_i\right)^{9/2}}.
\label{eq:final_PSD}
\end{equation}
\end{widetext}
The aim of these calculations is to find the value of $\rho_f (0)$ in the quadrupole magnetic trap and the value of field gradient $b$ which results in maximum value of $\rho_f (0)$, when the quadrupole magnetic field is switched-on to trap the atom cloud from the molasses at temperature $T_i$ and size $\sigma_i$. \figref{fig:theory} shows the calculated variation in temperature $T_f$ (given by equation \eqref{eq:Tf}), the final peak number density $n_f^\prime (0)$ (given by equation \eqref{eq:final_ND}) and the final peak phase-space density $\rho_f(0)$ (given by equation \eqref{eq:final_PSD}) in the magnetic trap with quadrupole field gradient $b$, for different values of temperature $T_i$ and size $\sigma_i$ of the atom cloud in the molasses.

With increase in the $b$, $\rho_f(0)$ first increases and then decreases after attaining a maximum value at a certain field gradient $b$. The reduction in $\rho_f(0)$ after the maximum value is due to increase in the temperature of cloud in the magnetic trap and saturation in number density with $b$. The temperature ($T_f$) increases with $b$ due to increase in potential energy gained by the cloud from the trap. For lower values of $b$, the number density in the magnetic trap is low which results in lower values of $\rho_f(0)$. Therefore a maximum for $\rho_f(0)$ is obtained in its variation with the field gradient $b$. 

We also note from \figref{fig:theory}, that a molasses cloud at higher temperature and smaller size ($T_i=100~\mu K$ and $\sigma_i=0.55~mm$) can lead to a higher final phase-space density ($\rho_f(0)$) in the magnetic trap than that obtained with a molasses cloud of lower initial temperature and larger size ($T_i=70~\mu K$ and $\sigma_i=0.70~mm$). This is because the smaller value of $\sigma_i$ can result a lower final temperature in the magnetic trap $T_f$, as shown in \figref{fig:theory}[a].

\section{Experimental}
\label{sec:experimental}

Experiments have been performed to verify the theoretical predictions as discussed in the previous section. The magnetic trapping experiments have been performed in a double magneto-optical trap (double-MOT) setup developed for BEC of $^{87}$Rb atoms. The schematic diagram of the setup is as shown in \figref{fig:dMOT_&_pulses}(a) and more details are described in ref. \cite{Ram2013a}. This setup consists of a vapor cell MOT (VC-MOT) which is formed in a chamber at $\sim1\times10^{-8}$ Torr pressure (with Rb-vapor), and an ultra-high vacuum MOT (UHV-MOT) which is formed in a glass cell at $\sim6\times10^{-11}$ Torr pressure. The UHV-MOT is loaded by transferring atoms from the VC-MOT using a red-detuned push laser beam. The atom cloud in the UHV-MOT is used for the magnetic trapping as UHV environment in the glass cell suits well for a long life-time of atoms in the magnetic trap. To transfer atoms from VC-MOT to UHV-MOT, a red detuned (detuning value $\delta/2\pi = -1.0~ GHz$ with respect to peak of $\cool$ transition of $^{87}$Rb atom) push laser beam is focused on the VC-MOT. The duration and sequence of various stages from VC-MOT formation to magnetic trapping and detection are shown schematically in \figref{fig:dMOT_&_pulses}(b). Typically, the number of $^{87}$Rb atoms which can be obtained in VC-MOT and UHV-MOT in this setup are $\sim1\times10^{8}$ and $\sim2\times10^{8}$ respectively after setting the appropriate values of various parameters in the setup. Atoms in the UHV-MOT are trapped in the magnetic trap after they are cooled in the compressed-MOT and molasses stages. The atoms in the UHV-MOT are kept in a compressed UHV-MOT for $\sim20~ ms$ duration, which is implemented by increasing the detuning of the UHV-MOT cooling laser beams to the value $\sim50~ MHz$ to the red side of the cooling transition peak. The compressed MOT implementation leads to lowering of the temperature as well as increase in the density in the atom cloud in the UHV-MOT, which is useful for loading the magnetic trap. After the compressed UHV-MOT, these cold atoms are kept in the molasses for a variable time duration ($3-9~ ms$) to further lower the temperature of the atom cloud for magnetic trapping. The atoms cooled in the optical molasses are then optically pumped to ($5S_{1/2} ~\left|F = 2,~ m_F = 2\right\rangle$) state for trapping in the quadrupole magnetic trap. The sequence of various stages shown in \figref{fig:dMOT_&_pulses}(b) is experimentally accomplished using different electronic pulses generated from a Controller system to control several acousto-optic modulators (AOMs), power supplies and switching circuits. This Controller system is operated through a PC (Personal Computer) and LabVIEW program. The electronic pulses from the Controller system are also used to generate probe laser pulses (using AOMs) and to trigger the detectors such as CCD camera for the characterization of cold atom cloud in the MOTs and magnetic trap. 

\begin{figure}[!h]%
\centering
\subfigure[]{\includegraphics[width=0.9\columnwidth]{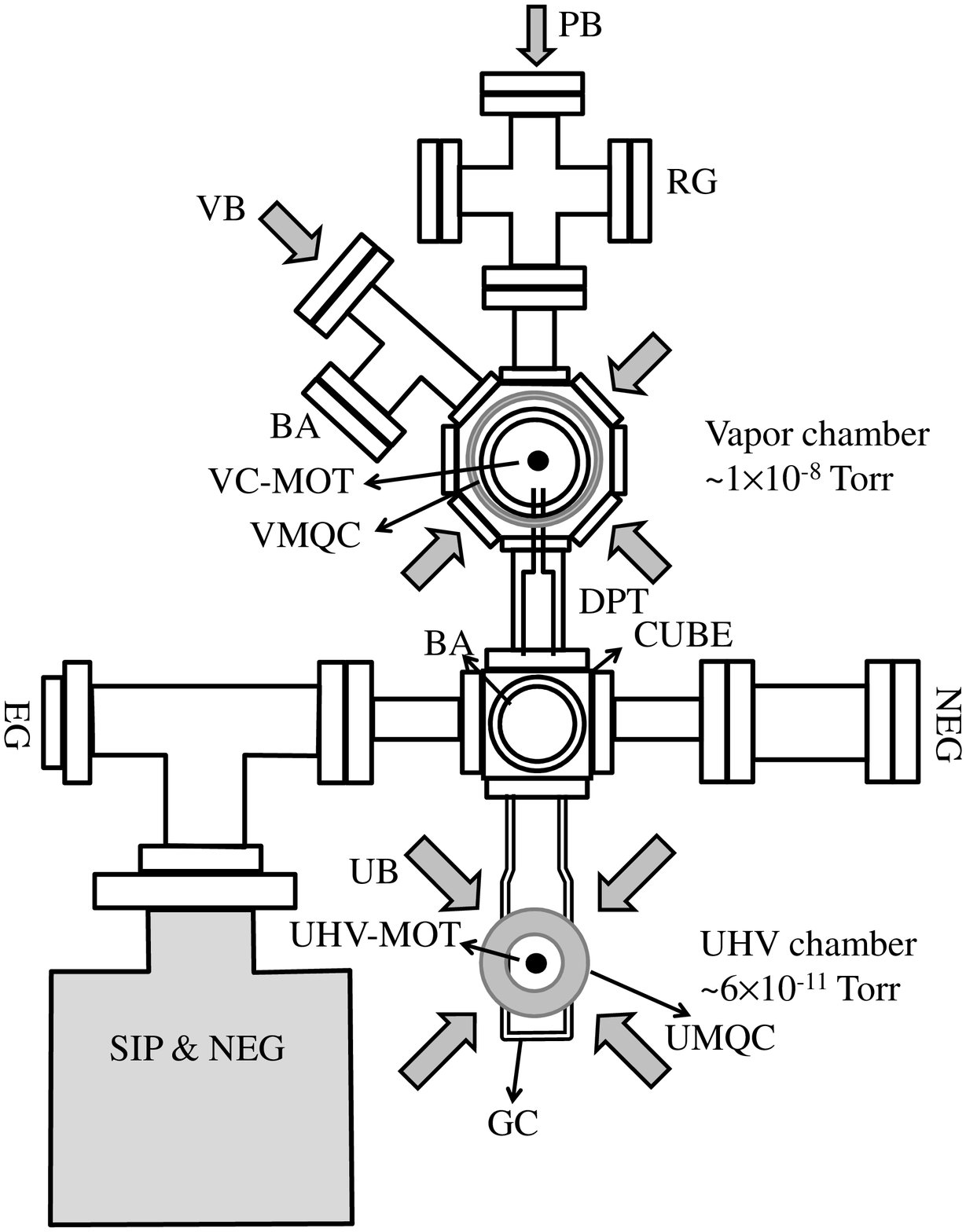}}
\subfigure[]{\includegraphics[width=0.9\columnwidth]{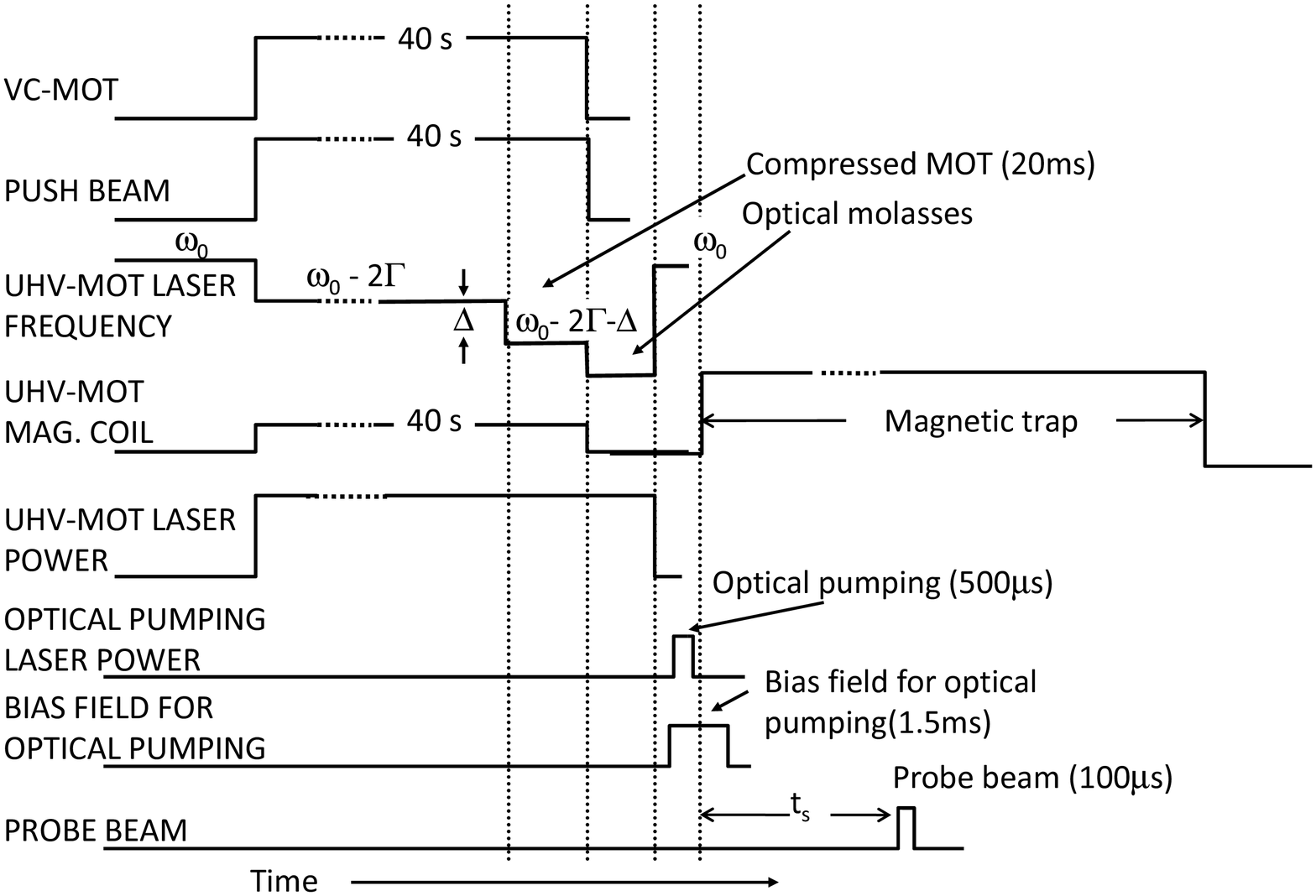}}
\label{fig:pulse_sequence}
\caption{(a) The schematic of the experimental setup. VB: VC-MOT beams, UB: UHV-MOT beams, PB: push beam, VMQC: VC-MOT quadrupole coil, UMQC: UHV-MOT quadrupole coil, RG: Rubidium getters, DPT: differential pumping tube, GC: glass cell, BA: Bayard-Alpert gauge, EG: extractor gauge, NEG: non-evaporable getter pump, SIP \& NEG: combination pump. (b) The sequence and durations of various stages from VC-MOT formation to magnetic trapping and detection of $^{87}$Rb atoms in the experiments.
}
\label{fig:dMOT_&_pulses}%
\end{figure}

To perform the optical pumping of atoms to ($5S_{1/2} ~\left|F = 2,~ m_F = 2\right\rangle$) state, the small parts of the cooling and re-pumping laser beams (with powers $\sim2~ mW$ in each part) are mixed and the combined beam is passed through an AOM in the double pass configuration. The output of this AOM ($\sim 500~ \mu W$, peak intensity of the beam is $\sim 1.6 ~mW/cm^2$), called optical pumping beam, is aligned to one of the UHV-MOT beam. The polarization of the optical pumping beam is made circular using a quarter wave-plate. This circularly polarized optical pumping beam (500 $\mu s$ duration) in presence of a small bias field ($\sim 2 ~G$, $\sim 1.5~ms$ duration) transfers the laser cooled atoms from molasses to state ($5S_{1/2} ~\left|F = 2,~ m_F = 2\right\rangle$). The optical pumping beam parameters (power and duration) and the bias field parameters (strength and duration) are varied and corresponding variation in number of atoms in the trap is recorded. Finally, these parameters are set to obtain the maximum number in the quadrupole magnetic trap. 

A pair of water-cooled quadrupole coils (UMQC in \figref{fig:dMOT_&_pulses}(a)) for UHV-MOT as well as for magnetic trapping has been used in the experiments. To switch-on the current in these quadrupole coils, an IGBT (Insulated Gate Bipolar transistor) based switching circuitry has been used which results in the current rise-time in the coils $\sim2.5 ~ms$. The current switch-off time of this circuitry is much shorter ($\sim 100~\mu s$) than the rise-time. The switching circuit receives the trigger pulse from the Controller (shown in \figref{fig:dMOT_&_pulses}(b)) to switch the current in the coils for UHV-MOT formation as well as for magnetic trapping. For magnetic trapping, a much higher value of current in these quadrupole coils is used in the experiments. The number of atoms in the magnetic trap has been estimated using the well known fluorescence imaging method \cite{SRMishra2008a} which uses a resonant probe laser pulse of short duration ($\sim100 ~\mu s$) to shine the trapped atom cloud and collect the emitted fluorescence on a CCD camera through an imaging optics. The temperature measurement for the atom cloud in the MOT or magnetic trap involves the similar imaging method, but the imaging is done during free expansion of the cloud \cite{Myrskog2000a} after its release from the MOT or magnetic trap. This free-expansion method has been used by us to estimate the temperature of cloud in the molasses and magnetic trap. The temperature of the atom cloud in the magnetic trap has also been measured by measuring the size of the cloud in the trap. The temperature values obtained by both these methods have shown a reasonable agreement, which is discussed in the next section of this paper.

\section{Results and Discussion}
\label{sec:results}
In the experimental studies, the temperature and number of atoms in the atom cloud in the quadrupole magnetic trap has been estimated only when atom cloud after molasses and optical pumping stages gets trapped in the quadrupole trap and reaches a nearly steady state equilibrium with the trap. \figref{fig:loading_MT} shows the measured temporal variation in the CCD counts (which is proportional to number of atoms) of the cloud image taken after collecting the probe induced fluorescence from atoms in the quadrupole magnetic trap. The time shown in the graph is the delay in recording the image of cloud after switching-on current in the quadrupole magnetic trap coils (current $\sim 13~ A$). It is evident from the figure that, after an initial sharp fall in the counts, the CCD counts becomes nearly constant for time duration $>50~ ms$. Thus the number of atoms measured after $\sim 50 ~ms$ duration corresponds to the number actually trapped in the magnetic trap. The sharp initial fall in \figref{fig:loading_MT} is due to fast removal of un-trappable atoms from the trap. The \figref{fig:T_MT}(a) which shows the temperature of the atoms in the magnetic trap as a function of time spent in the magnetic trap also reveals that after $\sim 50 ~ms$ duration, the temperature of atom cloud in the magnetic trap does not change considerably. These results thus suggest that, after the atoms have spent time $ >50 ~ms $ in the magnetic trap, the measurements of number, temperature and density of the atom cloud in the trap should be appropriate to characterize the atom cloud in the magnetic trap. 

 \begin{figure}[t]%
\centering
 \includegraphics[width=0.9\columnwidth]{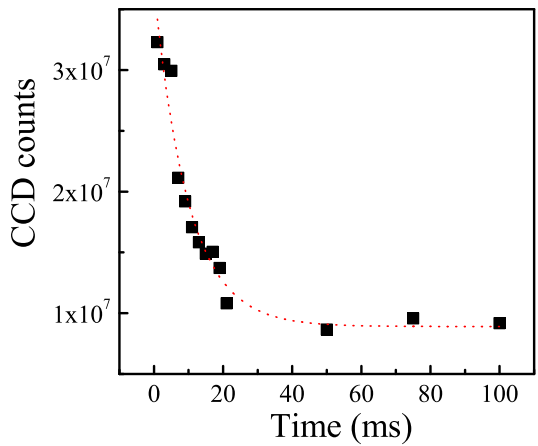}%
 \caption{The variation in the CCD counts due to fluorescence from atoms trapped in the quadrupole magnetic trap with time for $\sim 13~ A$ ($b= 95~G/cm$) of current in the quadrupole coils. The dotted curve is a guide to the eye.}%
 \label{fig:loading_MT}%
 \end{figure}

\begin{figure}%
\centering
\subfigure{\includegraphics[width=0.9\columnwidth]{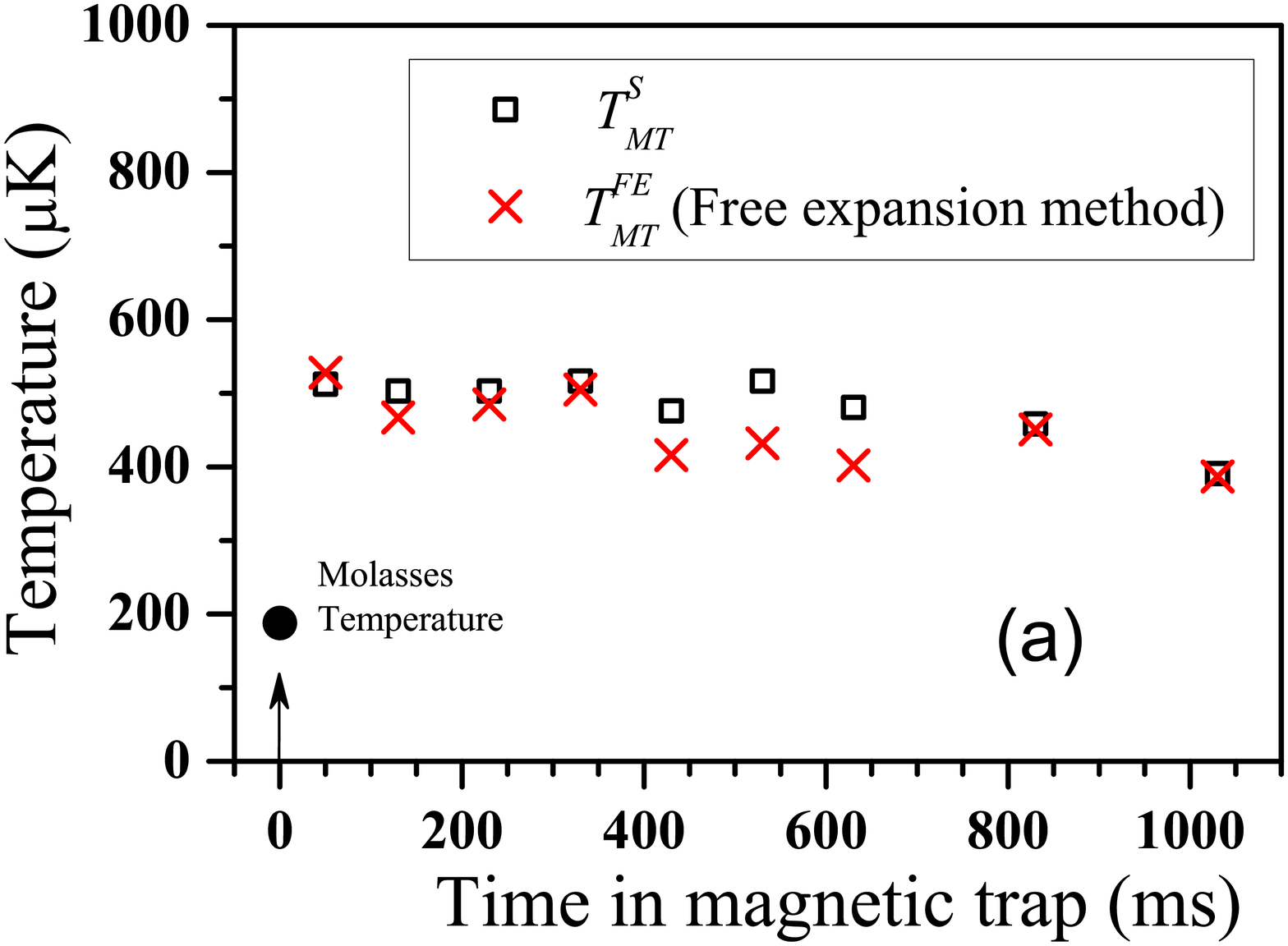}}%
\hfill
\subfigure{\includegraphics[width=0.9\columnwidth]{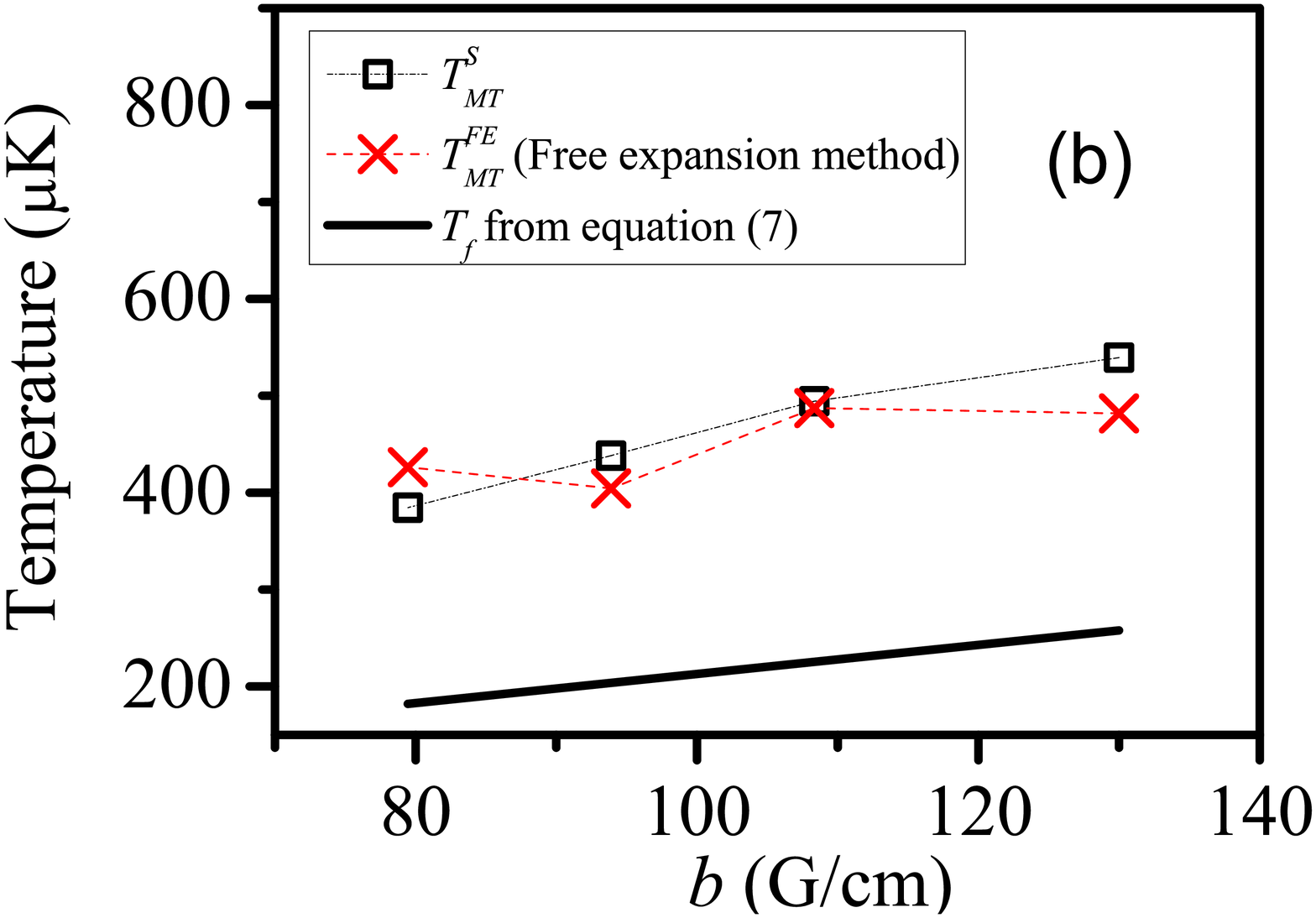}}%
\caption{(a): Measured variation in the temperature of atom cloud in the magnetic trap with the time duration in the magnetic trap for $\sim18~ A$ ($b$ = 130 $G/cm$) of current in the quadrupole coils. (b): Measured variation in the temperature of atom cloud in the magnetic trap with the field gradient $b$. In both the graphs squares show the temperature measured from the size of the cloud in magnetic trap, and crosses show the temperature measured by free expansion method. For these measurements the atom cloud temperature and size after the molasses were $T_i\sim$ 188 $\mu K$ and $\sigma_i\sim ~0.91~ mm$ respectively.}%
\label{fig:T_MT}%
\end{figure}

The measurement of temperature of atom cloud in the magnetic trap has been performed by two independent methods. The first method is based on the measurement of the full width at half maximum (FWHM) size ($R_{FWHM}$) of the atom cloud in the magnetic trap for a given field gradient $b$ and using the relation \cite{Lewandowski2003a, Yum2012a},

\begin{equation}
T_{MT}^S = \frac{2}{5} \frac{\mu_B}{k_B} bR_{FWHM}.
\label{eq:T_MT}
\end{equation}
where, $T_{MT}^S$ denotes the temperature of atom cloud in the magnetic trap measured by the size of the cloud. The temperature measured by this method is shown by squares in \figref{fig:T_MT}. The second method which we used is free-expansion method \cite{Myrskog2000a}. In this method the images of the atom cloud released from the magnetic trap are recorded at different times during the expansion of the cloud, and temperature ($T^{FE}_{MT}$) is estimated from the rate of change in the size of the cloud. The temperature measured by this method is shown by crosses in \figref{fig:T_MT}. Both these temperature measurement methods were tested on a cloud having the same initial temperature and size values after the molasses ($T_i\sim$ 188~ $\mu K$ and $\sigma_i\sim 0.91~mm$). The data in the figure shows that temperature obtained from the two methods have shown a reasonable agreement. \figref{fig:T_MT}(a) shows the variation in temperature of the atom cloud in the magnetic trap with time whereas \figref{fig:T_MT}(b) shows the variation in the temperature of the atom cloud in the magnetic trap with field gradient $b$. \figref{fig:T_MT}(a) shows that temperature does not change considerably with time spent in the magnetic trap. However, from \figref{fig:T_MT}(b), it can be noted that the measured temperature of the cloud in magnetic trap is higher than the temperature expected from the equation \eqref{eq:Tf} of theory (which is shown by a straight line in the figure \figref{fig:T_MT}(b)). These results are similar to the results of Stuhler et al. \cite{Stuhler2001a}, who also have reported the measured temperature in magnetic trap higher than the theoretically expected temperature. This difference between experimentally measured and theoretically expected values of temperature could be due to several reasons. One of these could be the mismatch between the center of the atom cloud from molasses and the center of the magnetic trap. The increase in the temperature in the magnetic trap is expected to lower the phase-space density of the cloud.

\begin{figure}%
\centering
\includegraphics[width=0.9\columnwidth]{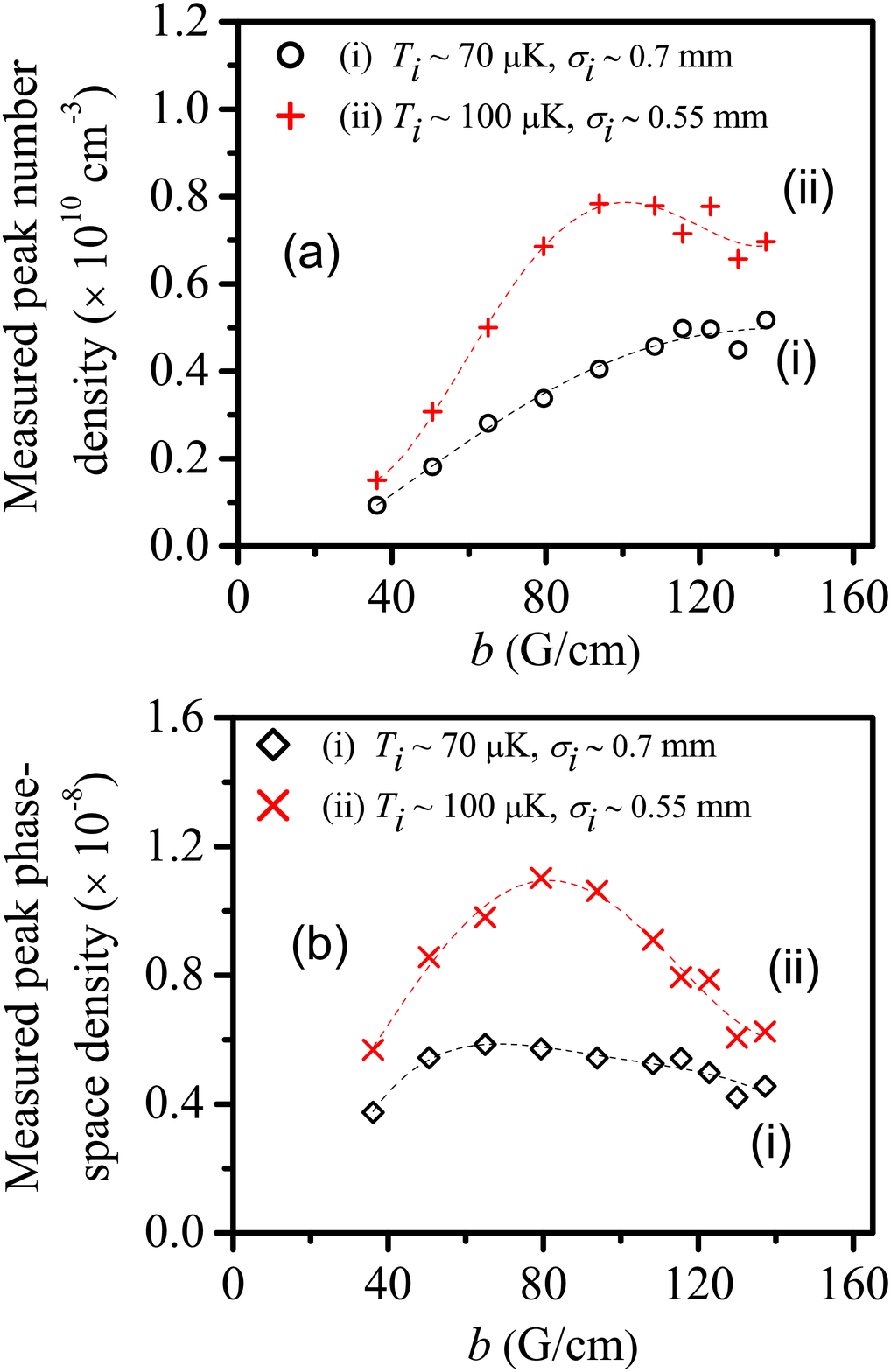}
\caption{Measured variation in the peak number density (a) and phase-space density (b) in the magnetic trap with the quadrupole field gradient $b$, for two sets of temperature and r.m.s. size of the cloud after molasses. The number of atoms ($N^\prime$) in the magnetic trap also varied with $b$. The maximum values of $N^\prime$ were $\sim 3.0\times 10^7$ and $\sim 5.5\times 10^7$ for graphs (i) and (ii) in this figure. The dashed curves are guide to the eye.}%
\label{fig:ND_PSD_expt}%
\end{figure}

The \figref{fig:ND_PSD_expt} shows the measured variation in the number density and phase-space density of the atom cloud in the magnetic trap with the quadrupole field gradient $b$, for different values of initial temperature ($T_i$) and size ($\sigma_i$) of the cloud in the molasses. Here peak number density has been estimated by measuring the number of atoms and size of atom cloud in the magnetic trap. These data are obtained from the fluorescence image of the trapped cloud. The temperature was obtained from the size of the cloud (using equation \eqref{eq:T_MT}) to know the peak phase-space density of the trapped cloud shown in \figref{fig:ND_PSD_expt}. It is evident from these results that measured values of phase-space density in the magnetic trap also first increases with $b$ and then decreases with it, after attaining a maximum value at an optimum $b$. Thus measured number density and phase-space density have shown the similar trend in variation with $b$ as predicted by theory (\figref{fig:theory}), for two different values of $\sigma_i$ and $T_i$. However, as one can note from data in \figref{fig:theory} and \figref{fig:ND_PSD_expt}, the measured values of number density and phase-space density in the magnetic trap are significantly lower than their theoretically predicted values. We attribute this difference between measured and theoretical values of both the parameters ($n$ and $\rho$) to the higher value of the measured temperature than the theoretically predicted temperature. Nevertheless, there exists an optimum field gradient at which the phase-space density in the magnetic trap is experimentally maximized.

\begin{figure}%
\centering
\includegraphics[width=0.9\columnwidth]{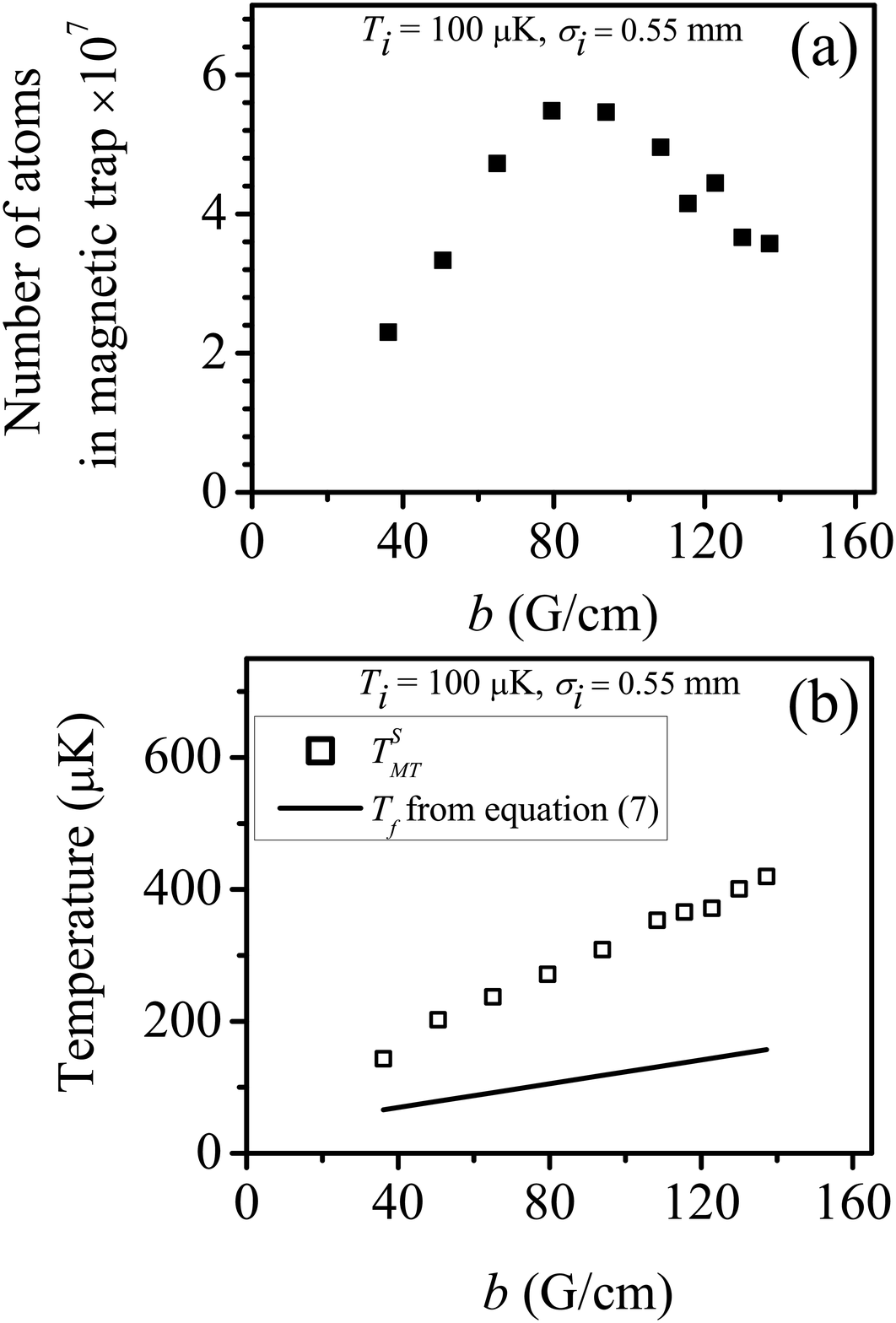}
\caption{(a): The measured variation in the number of atoms ($N^\prime$) trapped in the magnetic trap with the magnetic field gradient $b$. (b): The variation in measured temperature and the theoretically expected temperature of atom cloud in the magnetic trap with the magnetic field gradient $b$. The temperature and size after the molasses were $T_i\sim$ 100~ $\mu K$ and $\sigma_i\sim 0.55~mm$.}%
\label{fig:N_vs_b}%
\end{figure}

In the theory, equations \eqref{eq:final_ND} and \eqref{eq:final_PSD}, we have assumed that number actually trapped in the magnetic trap $N^\prime$ is independent of field gradient $b$. But actually $N^\prime$ changes with $b$, since $\varepsilon$ depends upon the gradient $b$ and switching time of the magnetic trap to reach the set $b$ value. This is due to escape of more energetic atoms from the trap for low values of $b$. The variation of measured $N^\prime$ with the magnetic field gradient $b$ is shown in \figref{fig:N_vs_b}(a). However, this variation of $N^\prime$ with $b$ (\figref{fig:N_vs_b}(a)) can not account for the difference between theoretical and experimental values of phase-space density (as well as number density). Our temperature measurement data support that the pronounced difference between theoretical and experimental values of phase-space density (also number density) is due to difference in the theoretical and experimentally measured temperature of atom cloud in the magnetic trap (\figref{fig:N_vs_b}(b)). This is because, both, number density and phase-space density, are highly non-linear functions of the temperature. 

\section{Conclusion}
\label{sec:conclusion}
We have studied the modifications in the temperature, number density and phase-space density of the cold atom cloud when it is transferred from the molasses to a quadrupole magnetic trap. It has been shown theoretically that, for a given temperature and size of the atom cloud in the molasses, there is an optimum value of the magnetic field gradient which results in the maximum phase-space density of the atom cloud in the quadrupole magnetic trap, when molasses cloud is trapped in the quadrupole magnetic trap. The experimentally measured variation in the phase-space density in the magnetic trap with the magnetic field gradient has shown the similar trend, with a maximum value of phase-space density observed at an optimum field gradient. However, the experimentally measured values of number density and phase-space density are much lower than their theoretically predicted values. This has been attributed to the higher values of experimentally observed temperature in the magnetic trap than the theoretically predicted temperature. These results, however, guide us to choose the appropriate value of the magnetic field gradient to be switched-on for magnetic trapping of the cold atom cloud from the molasses. Further, the results of this study may guide one to set a higher initial phase-space density of the cloud in the quadrupole magnetic trap, which may be useful when this trap is converted into a hybrid trap or a QUIC trap for the evaporative cooling to achieve BEC.

\begin{acknowledgments}

We thank our colleagues V. B. Tiwari and Surendra Singh for their helpful suggestions on the manuscript. We thank Lalita Jain, V. P. Bhanage, P. P. Deshpande, M. A. Ansari, H. R. Bundel and C. P. Navathe (all are from LESD, RRCAT) for development of the Controller system for the setup and H. S. Vora, (LESD, RRCAT) for providing the image-processing software. We also thank K. V. A. N. P. S. Kumar (UHVT Lab, RRCAT) and S. K. Shukla (ex-Head, UHVTD, RRCAT) for their help in the vacuum system.
\end{acknowledgments}


\end{document}